\newcommand{\logg}{$\log\,g$}
\newcommand{\teff}{$T_{\rm eff}$}

\documentclass{aa}
\usepackage{txfonts}
\usepackage{graphicx}
\begin{document}

\title{
A refined analysis of the remarkable Bp star HR\,6000 \thanks{Based
on observations collected at the European Southern Observatory, Paranal, Chile
(ESO programme 076.D-0169(A)).}}

\author{
F.\, Castelli
\inst{1}
\and
S.\, Hubrig
\inst{2}
}

\offprints{F. Castelli}

\institute{
Istituto Nazionale di Astrofisica--
Osservatorio Astronomico di Trieste, Via Tiepolo 11,
I-34131 Trieste, Italy\\
\email{castelli@oats.inaf.it}
\and
European Southern Observatory, Casilla 19001, Santiago 19, Chile\\
\email{shubrig@eso.org}
}

\date{}

\abstract
{}
{UVES spectra  of the very young ( $\sim$\,10$^{7}$ years) peculiar B-type star HR\,6000 were analyzed 
in the  near-UV and visual spectral regions (3050-9460\,\AA{}) with the aim to extend
to other spectral ranges the study  made previously in the UV using IUE spectra.} 
{
Stellar parameters \teff=12850\,K, \logg=4.10, and $\xi$=0\,km\,s$^{-1}$,
as determined from H$_{\beta}$, H$_{\gamma}$, H$_{\delta}$ Balmer profiles and  
from the \ion{Fe}{i}, \ion{Fe}{ii} ionization equilibrium, were used to compute 
an individual abundances ATLAS12 model. We identified spectral peculiarities 
and obtained final stellar abundances by comparing  observed and computed equivalent
widths and line profiles. 
}
{
The adopted model fails to reproduce the (b-y) and c color indices.
The spectral analysis has revealed: the
presence of emission lines for \ion{Mn}{ii}, \ion{Cr}{ii}, and \ion{Fe}{ii};  
isotopic anomalies for \ion{Hg}, \ion{Ca}; 
the presence of interstellar lines of \ion{Na}{i} at 
$\lambda\lambda$ 3302.3, 3302.9, 5890, 5896\,\AA{}, and of \ion{K}{i}
at 7665, 7699\,\AA{}; the presence of a huge quantity of unidentified lines,
which we presume to be mostly due to \ion{Fe}{ii} transitions owing to the
large Fe overabundance amounting to [+0.7].  
The main chemical peculiarities are an extreme overabundance of Xe, followed
by those of Hg, P, Y, Mn, Fe, Be, and Ti. The most underabundant element
is Si, followed by  C, N, Al, S, Mg, V, Sr, Co, Cl, Sc, and Ni. The silicon
underabundance [$-$2.9] is the lowest value for Si ever observed in any HgMn star.  
The observed lines of \ion{He}{i}
can not be reproduced by a single value of the He abundance, but they require
values ranging from [$-$0.8] to [$-$1.6]. Furthermore, when the observed and
computed wings of \ion{He}{i} lines are fitted, the 
observed line cores are much weaker than the computed ones.
From the present analysis we infer the presence of  vertical abundance stratification for He, Mn,
and possibly also P.
}
{}


\maketitle{}

\section{Introduction}

HR\,6000 (HD\,144667) is one of the most remarkable chemically peculiar (CP) stars. It does 
not fit any of the CP subclasses, but it seems to combine abundance anomalies
from a variety of Bp sub-types. It forms with the star HR\,5999 (HD\,144668)
the common proper motion visual binary system $\Delta$199 or
Dunlop\,199 (Bessell \& Eggen, 1972). 
The angular separation between the HR\,6000, which is the brighter more massive component,
and the secondary component, the well-known Herbig Ae star HR\,5999, is about 45\arcsec{}.
 
The main interest to study the chemical composition of  HR\,6000 
comes from the generally poor understanding of the occurrence of 
abundance anomalies in such a young object
with an estimated age of the order of 10$^{7}$ years.  In fact,
the $\Delta$\,199 double system is located close to the center of the Lupus\,3 
molecular cloud which is populated by numerous T\,Tauri stars. 
This has led to an assumption that this system has the same age as the cloud,
which is estimated to be (9.1$\pm$2.1)$\times$10$^{6}$ years (James et al., 2006).
 However, after the Hipparcos mission, the membership
of HR\,5999 and HR\,6000 to the cloud became rather questionable due to the 
uncertainties in the distance determination for the Lupus cloud.
The distance of HR\,5999 and HR\,6000 measured by Hipparcos
are 208$\pm$38\,pc and 240$\pm$48\,pc, respectively,
while the Lupus cloud distance is 150$\pm$10\,pc according
to Crawford (2000). Comer\`on et al.\ (2003) assigned a distance of 
about 200\,pc to the Lupus cloud, but this determination was made
by assuming a priori that $\Delta$199 belongs to the cloud. 
 
A ROSAT survey of Herbig Ae/Be stars presented by
Zinnecker \& Preibisch (1994) has detected strong
X-ray emission coming from the direction of HR\,6000.
To explain the X-ray origin,
a possible T\,Tauri companion for HR\,6000 was postulated by van den Ancker
et al.\ (1996) on the basis of an infrared excess in the energy distribution
relative
to predictions made for an effective temperature of \teff=14000\,K.
Siebenmorgen et al.\ (2000) fitted very short spectral regions of observed
flux in the mid IR to a black body of 13000\,K and noticed that the observed 
spectrum was featureless.

Neither spectrum nor radial velocity variability was found 
for HR\,6000 by Andersen \& Jaschek (1984), who
studied optical spectra taken at different epochs. 
However, van den Ancker et al.\ (1996) discovered long-term
variations  in the u, v, and b Str\"omgren magnitudes
with approximate amplitude of 0\fm{}03$\pm$0\fm{}01 in u, 
0\fm{}02$\pm$0\fm{}01 in v, and 0\fm{}01 in b. No variations in y
were found.  Kurtz \& Marang (1995) discovered  variations of 0.008 mag
in V with a period near to 2\,d and suggested that this period could be possibly 
caused by rotational variation in a spotted magnetic star.
This would imply a pole on orientation for
HR\,6000 as they measured $v \sin i$$\le$5~km~s$^{-1}$.

Catanzaro et al.\ (2004)
were the first ones who provided abundances from the optical range.
Previously,  optical spectra were studied only by 
Andersen \& Jaschek (1984) and Andersen et al.\ (1984) who identified 
the spectra from 3323\,\AA{} to 5317\,\AA{} and discussed
the abundances on the basis of the line intensities.
It is remarkable that the stellar parameters derived by
Catanzaro et al.\ (2004) from the Balmer profiles 
(\teff=12950$\pm$50~K, \logg=4.05$\pm$0.01) are 1000\,K lower
than those previously adopted and deduced from the photometry by
Castelli et al.\ (1985) (\teff=14000\,K, \logg=4.0),  by Smith (1997)
(\teff=13990\,K, \logg=4.29) and by van den Ancker et al.\ (1996) 
(\teff=14000\,K, \logg=4.3).

In this paper we examine the whole optical spectrum of HR\,6000 from
3050\,\AA{} to 9460\,\AA{} with the aim to extend 
the analysis performed on IUE spectra by Castelli et al.\ (1985) to the visible region and
to investigate the possible contamination of the stellar spectrum by a
close T\,Tau companion. 
We re-examine the parameter determination, the line 
identification, and the abundances. We also searched for the presence 
of peculiarities like emission lines and isotopic anomalies which were 
recently discovered in HgMn stars. 
Owing to the availability of spectra taken at different 
epochs we also investigated possible spectral variabilities.

The comparison of the observed and computed spectra as described in this
paper is available at the web-address given in the 
footnote\footnote{http://wwwuser.oat.ts.astro.it/castelli/hr6000/hr6000.html}.

\section{Observations and Radial Velocity}

Spectra of HR\,6000 were recorded at ESO on 2005 September 19 
(JD 2453632.53881) and on  2006 March 25 (JD 2453819.90226).
In both epochs we used the UVES DIC1 and DIC2 standard settings 
to cover the spectral 
range from 3050\,\AA{} to 9500\,\AA{}.
The slit width was set to $0\farcs{}3$ for the red arm, 
corresponding to a resolving power of 
$\lambda{}/\Delta{}\lambda{} \approx 110000$. For the blue arm, we used 
a slit width of $0\farcs{}4$ to achieve a resolving power of 
$\approx 80000$.
The spectra were reduced by the UVES pipeline Data Reduction Software (version 2.5; Ballester 
et al.\ 2000) and using standard IRAF routines. 
The signal-to-noise ratio (S/N) of the UVES spectra is very high, 
ranging from 150 in the near UV (3300\,\AA{}) to about 350 and 300 at 5000\,\AA{}
and 7100\,\AA, respectively.

There are two gaps in the observed range at $\lambda\lambda$ 4523--4769\,\AA{}
and 7536--7660\,\AA{}, which are caused by the physical gap between the two 
detector chips of the red CCD mosaic. 
In addition, the reduction in the spectral range 8076--8093\,\AA{} is affected by the presence of 
a bad column on the MIT CCD.

The heliocentric radial velocities derived from spectra observed on September 2005 
and on March 2006 are 2.6\,km\,sec$^{-1}$ and 1.4\,km\,sec$^{-1}$, respectively.
These values are consistent with the mean heliocentric radial velocity 
+2.6$\pm$1.8~\,km\,s$^{-1}$ of the Lupus cloud (James et al., 2006),
so that they strenghten the assumption of the membership of HR\,6000 to the Lupus cloud.
However, Catanzaro et al.\ (2004) measured 0.67$\pm$0.38\,km\,s$^{-1}$, while
Andersen \& Jaschek (1984) determined
$-$1.5$\pm$0.5\,km\,s$^{-1}$. In their study, the values obtained from different spectra
range from +0.6$\pm$0.8\,km\,s$^{-1}$ to $-$3.3$\pm$0.8\,km\,s$^{-1}$.
Such differences in measured radial velocities by different authors indicate that small radial 
velocity variations in HR\,6000 are possibly real.

The spectra were normalized to the continuum by using a modified version 
of the interactive code NORMA written by Bonifacio (1989).
In the original procedure  smoothed interpolated curves through selected
continuum points are obtained by means of Hermite spline functions
(subroutine INTEP by Hill (1982)). We replaced the spline functions
with a linear interpolation because in some other previous stellar
analyses performed by one of us (F.C.) it has given a continuum more close 
to that we would have drawn by hand. 
However, we checked that in the case of HR\,6000 the continua from 
the two different interpolations do not differ in a significant way. 
 When needed, the continuum was  adjusted over
6\,\AA{} intervals with the help of the synthetic spectrum.
 The continuous level
just longward of the Balmer and Paschen discontinuities is highly uncertain
because it is difficult to know where to put it. Also
the continuum drawn over the broad hydrogen lines is uncertain owing to the
not fully corrected distorsions of the echelle spectra.

\section{Stellar parameters}

Table\,1 summarizes the stellar parameters as determined from 
Str\"omgren, UBV, and near-infrared IJ photometry.
The observed  Str\"omgren and UBV indices
were taken from the 
Hauck \& Mermilliod (1998) Catalogue\footnote{http://www.unige.ch/gcpd/gcpd.html}, while
(V$-$I$_{c}$)=$-$0.05 was taken from the Hipparcos Catalogue
and (V$-$J)=$-$0.032 
was obtained from the J magnitude given by Th\'e et al.\ (1996). 
The  parameter determination is the result of an interpolation
of the observed dereddened indices in the grid of  
synthetic indices based on the NEWODF model atmospheres
(Castelli \& Kurucz, 2004) computed for metallicity [M/H]=0.0 and 
microturbulent velocity $\xi$=0.0\,km\,s$^{-1}$. The color grids
are available at the web-address given in the 
footnote\footnote{http://wwwuser.oat.ts.astro.it/castelli/colors.html}. 
For the Str\"omgren and UBV photometry the adopted 
grids were uvbybeta and UBV, while for the (V$-$I$_{c}$) and  
(V$-$J) indices the adopted grid was UBVRIJHKL.

Results from the Str\"omgren photometry are listed in the first 
panel of Table\,1. For E(B$-$V)=0.06, the parameters from the c$_{0}$ and $\beta$ indices 
are \teff=13827$\pm$200\,K, \logg=4.27$\pm$0.05.  
The reddening E(B$-$V)=0.06 was determined by Bessell \& Eggen  (1972)
and later confirmed by Th\'e \& Tjin A Djie (1978).
An uncertainty in E(B$-$V) of 0.02\,mag  corresponds to a 
difference in \teff\ of only 15\,K. 
The reddening  E(B-V)=0.042 listed in the first row of Table\,1 corresponds to  
E(b$-$y)=0.031 as yielded by the UBVYLIST code of Moon (1985) on the basis
of standard reddening relations. The conversion 
 E(b$-$y)=0.74E(B$-$V) from Crawford \& Mendwewala (1976) was applied
here.

A microturbulent velocity of 2.0\,km\,s$^{-1}$ lowers the temperature
by 100\,K and does not affect the gravity, while an increase in metallicity
to [M/H]=+0.5 decreases \teff{} by about 50\,K; a decrease in
metallicity to [M/H]=$-$1.0 increases \teff{} by about 
+300\,K. The corresponding differences in \logg\ are of 
the order of 0.01\,dex in all the cases. 
These values are only indicative of
the metallicity effect in that 
the abundance pattern of HR\,6000  is very different from any scaled solar one.

The parameters  from the UBV photometry and
$\beta$ index are listed in the second panel of Table\,1.
For E(B-V)=0.06, [M/H]=0.0, and $\xi$=0.0 km\,s$^{-1}$ 
they are \teff{}=13973$\pm$300\,K, \logg=4.38$\pm$0.2 (Fig.\,1).
They are close to those obtained from the Str\"omgren photometry.
This parameter determination, described in Castelli et al.\ (1985),
is very sensitive to small reddening changes. 
In fact, for E(B-V)=0.04 or E(B-V)=0.08  there is no intersection 
in the curves (U-B)$_{0}$, (B-V)$_{0}$ and
$\beta$. For B-type stars, 
the (U-B), (B-V), and $\beta$ indices do not depend on the metallicity in
a significant way (Castelli, 1999; Castelli \& Kurucz, 2006).

\begin{figure}
\centering
\resizebox{3.5in}{!}{\rotatebox{90}{\includegraphics[0,170][550,750]{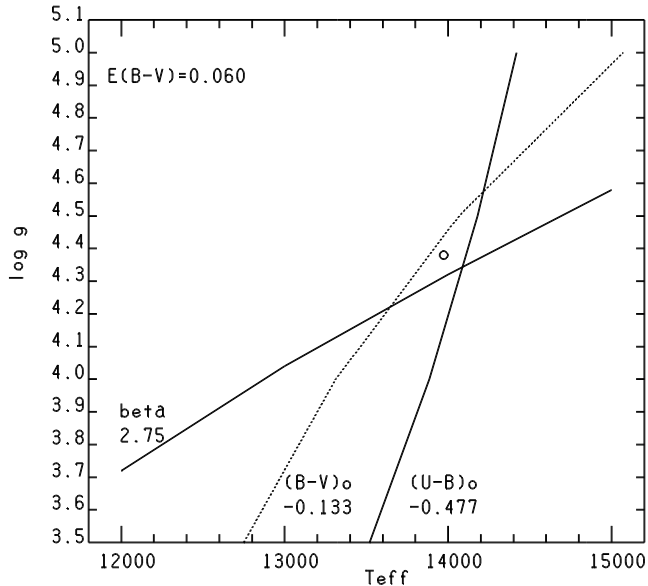}}}
\caption{Determination of the atmospheric parameters (\teff{},\logg) from (U$-$B)$_{0}$, (B$-$V)$_{0}$ and $\beta$. The circle represents the parameters
\teff{}=13873\,K, \logg=4.38 obtained from this method for [M/H]=0.0 and $\xi$=0.0 km\,s$^{-1}$.
The observed indices were dereddened for E(B$-$V)=0.060}
\label{fig1}
\end{figure}

The third panel of Table\,1 shows that 
for a prefixed gravity \logg=4.0, the observed (V-I)$_{c}$ and (V-J) indices 
are reproduced by  \teff{} equal to
10889\,K and 11517\,K, respectively, when 
[M/H]=0.0, $\xi$=0.0\,km\,s$^{-1}$, and E(B-V)=0.06. We used the
reddening relations E(V$-$I$_{c}$)=1.25E(B$-$V) from Dean et al.\ (1978) and
E(V$-$J)=2.23E(B$-$V) from Mathis (1990). 
The temperature was derived by interpolating the observed dereddened
(V$-$I$_{c}$)$_{0}$ and (V$-$J)$_{0}$ indices in the 
Bessell et al.\ (1998) color grid computed for [M/H]=0.0 and 
$\xi$=0.0~km~s$^{-1}$.

\begin{table*}
\caption{Model parameters for HR~6000 from photometry by assuming [M/H]=0.0 and $\xi$=0.0~km~s$^{-1}$.}
\centering
\begin{tabular}{llll|llll|lllll}
\hline
\hline\noalign{\smallskip}
\multicolumn{4}{c}{Observed indices}&&\multicolumn{3}{c}{Dereddened indices}&
\multicolumn{2}{c}{Parameters}\\
\hline\noalign{\smallskip}
 (b-y)    & m$_{1}$ & c$_{1}$ & $\beta$ & E(b-y)&(b-y)$_{0}$ &  m$_{0}$ &   c$_{0}$& \teff{}(K) &\logg&E(B-V) \\
\hline\noalign{\smallskip}
 $-$0.030 & 0.116   & 0.512   & 2.750   & 0.031 & $-$0.061   & 0.126    &  0.506   & 13799 &4.27 &0.042\\  
$\pm$ 0.003 &$\pm$0.003 &$\pm$0.003   &$\pm$ 0.013&       &            &          &       & $\pm$ 150  &$\pm$0.05     \\
          &         &         &         &0.044  & $-$0.074   & 0.130    &  0.503   &  13827 & 4.27&0.060\\
          &         &         &         &       &            &          &          &$\pm$200&$\pm$ 0.05\\
\hline\noalign{\smallskip}
\hline\noalign{\smallskip}
 V    & U-B &  B-V  & &E(B-V) &  V$_{0}$ & (U-B)$_{0}$ & (B-V)$_{0}$&  \teff{}(K) & \logg\\
\hline\noalign{\smallskip}
 6.647 & $-$0.434 &$-$0.073 && 0.06   &   6.461   &$-$0.477       &$-$0.133        &13973 &4.38\\
$\pm$0.013 &$\pm$ 0.004  &$\pm$ 0.004& &        &          &             &              &$\pm$300 &$\pm$0.2\\
\hline\noalign{\smallskip}
\hline\noalign{\smallskip}
      & (V-I)$_{c}$ & (V-J)   &&E(V-I$_{c}$) & E(V-J) &(V-I$_{c}$)$_{0}$& (V-J)$_{0}$&\teff(V$-$J)&\teff(V$-$I$_{c}$)&\logg \\
\hline\noalign{\smallskip}
      & $-$0.05      & $-$0.032&&0.075       & 0.134  &$-$0.125            & $-$0.166  &11517 & 10889&4.0 (prefixed)\\
      &$\pm$0.00\\
\hline\noalign{\smallskip}
\hline\noalign{\smallskip}
\multicolumn{3}{l} {model}  &                 &                         &(b-y)$_{calc}$&        & c$_{calc}$&\\
\hline\noalign{\smallskip}
\multicolumn{3}{l}{T12950G405k2M05nohe}&                                 &       &$-$0.041      &        & 0.629\\ 
\multicolumn{3}{l}{T12850g41k0at12}  &                                   &       &$-$0.049      &        & 0.596\\
\hline\noalign{\smallskip}
\end{tabular}
\end{table*}

Catanzaro et al.\ (2004) were the first ones who derived parameters
for HR\,6000 from high resolution spectra, in particular from
H$_{\gamma}$ and H$_{\delta}$ Balmer profiles observed on FEROS spectra.
They fixed \teff{}=12950$\pm$50\,K and \logg=4.05$\pm$0.01 from
ATLAS9 models computed for [M/H]=$-$0.5 and zero helium abundance.
This particular grid of models  yields the same parameters from both Balmer 
profiles, but we note that the parameters do not change if 
the grid  computed with [M/H]=0.0 and solar He abundance is used, 
provided that the different parameters
from H$_{\gamma}$ ([12750\,K,3.95] and H$_{\delta}$ ([13150\,K, 4.13])
are averaged.
As discussed in Sect.\,4, parameters very close to those of Catanzaro et
al. \ (2004) were derived by us from Balmer profiles and
\ion{Fe}{i}, \ion{Fe}{ii} equivalent widths measured on UVES spectra 
and from
model atmospheres computed for the individual abundances
of HR\,6000.

The last panel of Table\,1 shows that the cooler model,
as adopted by Catanzaro et al. (2004) 
(model T12950g405k2M05nohe), does not reproduce 
the observed Str\"omgren indices. In fact, the difference 
between the observed and computed (b$-$y)$_{0}$ indices amounts 
to  0.020\,mag, or even more, depending on the adopted reddening, 
while the difference  between the c$_{0}$ indices is 0.122\,mag. 
These discrepancies do not change in a remarkable way for the ATLAS12
model (model T12850g41k0at12) adopted by us to predict the observed
spectrum (see Sect.\,4).  On the other hand, the
hotter model [13900,4.32] fails to reproduce 
all the Balmer profiles. For solar abundances it 
provides good agreement  between observations and
computations for H$_{\alpha}$, but the agreement decreases 
from H$_{\beta}$ to H$_{\delta}$ in the sense that the computed wings
become too broad as compared with the observed ones; for individual abundances
the computed inner wings of all the profiles are narrower than 
the observed ones, in particular for H$_{\alpha}$ whose  profile seems also 
to be a little bit asymmetric with a more extended red wing. 
All the comparisons between
the observed and computed Balmer profiles are available at the 
website of footnote\,1. Fig.\,2 shows the case of H$_\beta$.

\begin{figure*}
\centering
\includegraphics[height=\textwidth,angle=90]{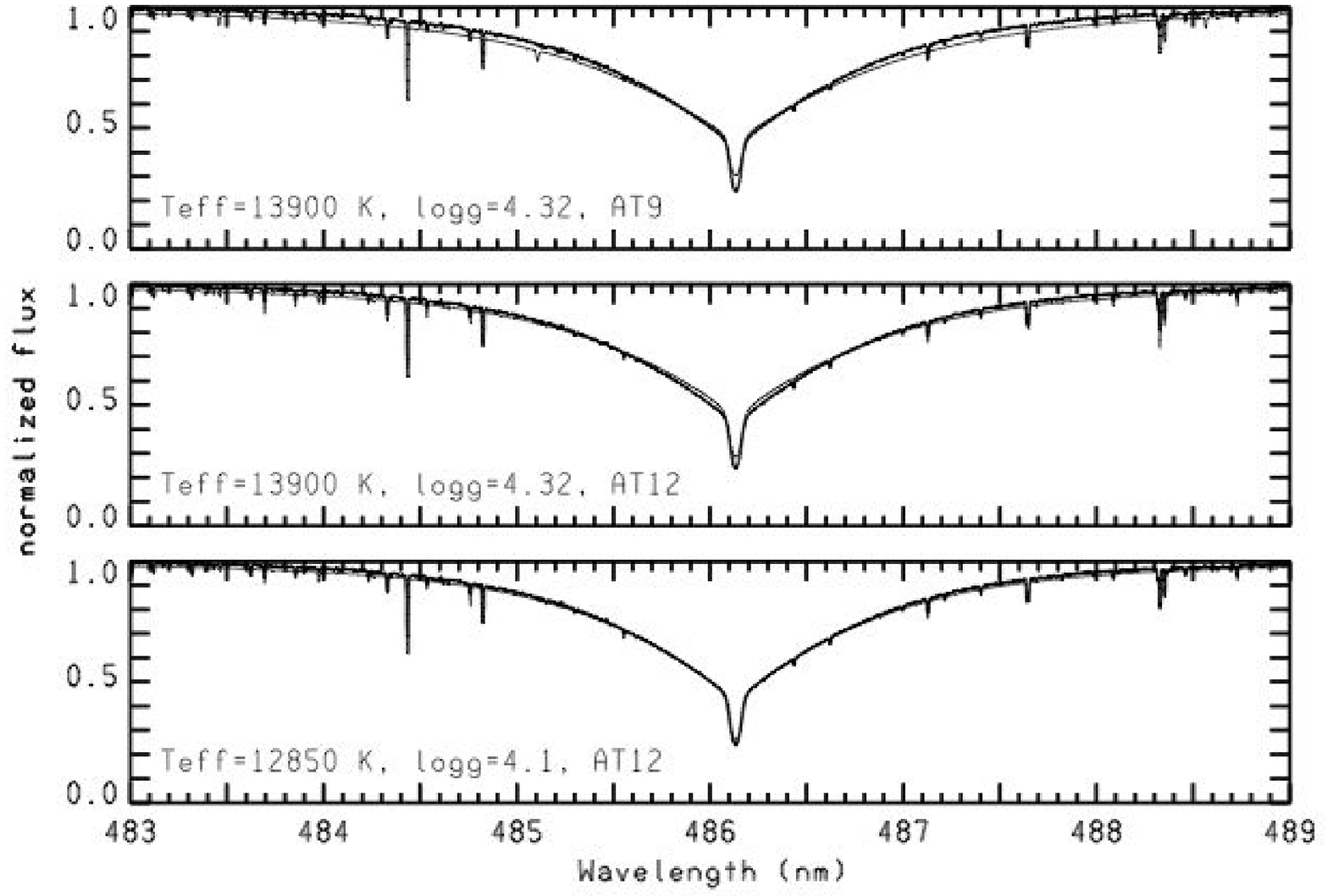}
\caption{The observed H$_{\beta}$ (thick line) is compared with  profiles
computed with different models (thin lines). In the upper and middle panels
the ATLAS9 model and the ATLAS12 model with parameters given by the photometry 
(\teff=13900\,K, \logg=4.32, $\xi$=0.0\,km\,s$^{-1}$) were used. 
The lower panel shows the profile 
computed with the final adopted ATLAS12 model with parameters \teff=12850\,K,
\logg=4.1, and $\xi$=0.0\,km\,s$^{-1}$.}
\label{fig2}
\end{figure*}

In conclusion,  there is a discrepancy of about 1000\,K between 
parameters derived from UBV and Str\"omgren photometry and 
parameters derived from the line spectrum. Furthermore, the
difference in \teff\ from the near-IR photometry is so
large that spectrophotometric observations are needed in order
to confirm these results. Unfortunately, the only  
spectrophotometric data available for HR\,6000 are the IUE spectra.
For E(B-V)=0.06, the best fit of the UV energy distribution 
from the IUE images SWP14849 and LWR11431 to the grid of ATLAS9 energy 
distributions computed for [M/H]=0.0 and $\xi$=0.0\,km\,s$^{-1}$ 
is achieved with \teff{}=13300\,K, \logg=4.1.
More details about these IUE images  can be found
in Th\'e et al.\ (1996).

\section{The ATLAS12 model and abundance determination } 

The opacity distribution function ATLAS9 model with 
parameters derived from
the UBV and Str\"omgren photometry (\teff{}=13900\,K, \logg=4.32, [M/H]=0.0,
$\xi$=0.0\,km\,s$^{-1}$) was used to derive preliminary abundances.
Abundances for iron and phosphorus were obtained from equivalent widths,
while those for all the other elements were obtained with 
the synthetic spectrum method.  Mostly
the lines listed in Castelli \& Hubrig  (2004b) (electronic Appendix)
were examined.  Synthetic spectra were computed 
with the  SYNTHE code (Kurucz, 1993). The adopted line lists are available
on the web-address given in the 
footnote\footnote{http://wwwuser.oat.ts.astro.it/castelli/linelists.html}.
The synthetic spectra
were broadened for a rotational velocity $v \sin i$=1.0\,km\,s$^{-1}$
and an instrumental resolving power of 80000 for $\lambda$$<$4520\,\AA{} 
and  of 110000 for $\lambda$$>$4765\,\AA{} (Sect.\,2).
The adopted $v \sin i$ was
that reproducing the observed profiles at best, in particular 
\ion{Mg}{ii} 4481\,\AA{}. It is in close 
agreement with the value of 0.0\,km\,s$^{-1}$ derived by
Catanzaro et al.\ (2004).

With this model, the average abundance derived 
from the equivalent widths of 11 \ion{Fe}{i} selected 
lines was equal, within the errors limits, to that obtained from 
the equivalent widths of 34 selected \ion{Fe}{ii} lines, i.e.
 $\log$(N(\ion{Fe}{i})/N$_{tot}$)=$-$3.65$\pm$0.06
and $\log$(N(\ion{Fe}{ii})/N$_{tot}$)=$-$3.69$\pm$0.15.
At the same time, for $\xi$=0.0\,km\,s$^{-1}$  there was no trend of
the abundances  versus the equivalent widths of the \ion{Fe}{ii} lines. 
The table with the measured equivalent widths  and
corresponding abundances is available at 
the website of footnote\,1.

The abundances based on the ATLAS9 model were used as input data
for an opacity sampling ATLAS12 model (Kurucz 1997;  Castelli \& Kurucz, 1994)
computed for the same parameters of the ATLAS9 model. 
Due to the extremely peculiar abundances of HR\,6000, the T-$\tau_{Ross}$ 
structure of the ATLAS12 model is rather different from that
of the ATLAS9 model, in particular 
in  layers upper than  $\log$($\tau_{Ross}$)$\le$ $-$0.5.
 The comparison of the T-$\tau_{Ross}$ relations from ATLAS9 and
ATLAS12 is similar to that shown in  Castelli \& Hubrig (2004b).
As a consequence, in spite of the achieved better agreement
between observed and computed Balmer profiles, 
the \ion{Fe}{i}-\ion{Fe}{ii} ionization equilibrium worsens. 
The average abundances  
for \ion{Fe}{i} and \ion{Fe}{ii} became $-$3.42$\pm$0.06\,dex and 
$-$3.61$\pm$0.15\,dex, respectively.

A search for an appropriate ATLAS12 model suitable to  reproduce both 
the Balmer profiles
and the \ion{Fe}{i}-\ion{Fe}{ii} ionization equilibrium  has finally yielded, after several
trials and errors, the parameters
\teff{}=12850~K, \logg=4.1, in good agreement with the values adopted 
by Catanzaro et al.\ (2004). The \ion{Fe}{i} and \ion{Fe}{ii} abundances were
$-$3.89$\pm$0.06\,dex and $-$3.88$\pm$0.16\,dex, respectively.
Fig.\,3 shows that this model also reproduces, 
for E(B-V)=0.06, the slope of the UV energy distribution given by
the IUE images SWP14849 and LWR11431. However,
a close inspection of Fig.\,3 reveals the presence of some
strong absorptions in the short-wavelength region which are neither
predicted by the models nor detected in the examined IUE high-resolution 
spectra (Sect.\,5.1). 
As already discussed in the previous section, the final ATLAS12 
model fails in reproducing the observed Str\"omgren indices.
The conclusion is that we were not able to fix a model well suited
to reproduce  the atmosphere of HR\,6000 as deduced
from the available observations.
Table\,2 summarizes the parameter values from the different determinations
based on the ATLAS9 (AT9) and ATLAS12 (AT12) models.

\begin{figure}
\centering
\includegraphics[height=0.5\textwidth,angle=90]{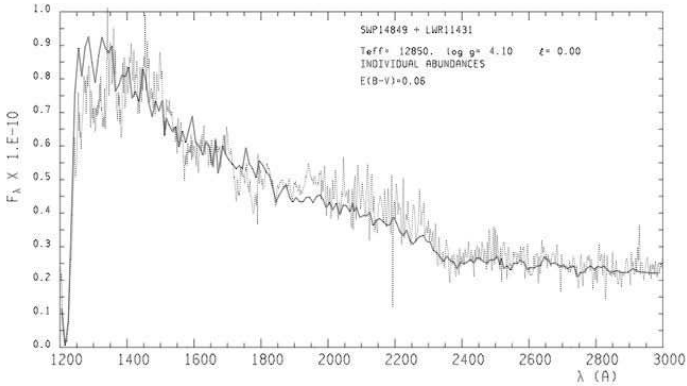}
\caption{Comparison of IUE data (dotted line) with the flux computed from 
the ATLAS12 model with individual stellar abundances (full line). 
Model parameters are \teff{}=12850\,K,  \logg=4.1, and $\xi$=0.0\,km\,s$^{-1}$. The IUE flux was dereddened 
for E(B-V)=0.06.}
\label{fig3}
\end{figure}

\begin{table}
\caption{Summary of the stellar parameters for HR\,6000 from different
determinations. }
\centering
\begin{tabular}{lllllllllllll}
\hline
\hline\noalign{\smallskip}
Method     & \teff{}  & \logg & Models       \\
\hline\noalign{\smallskip}
UBV$\beta$ & 13973$\pm$300 & 4.4$\pm$0.2 & AT9\\
c$_{0}$,$\beta$&13813$\pm$200&4.3$\pm$0.05 & AT9\\
IUE flux   &13300$\pm$100&4.1$\pm$0.1 & AT9 \\
Balmer profiles &12850$\pm$50 & 4.1$\pm$0.01&AT12\\
\ion{Fe}{i}/\ion{Fe}{ii} & 12850 & 4.1 &AT12\\ 
VI$_{c}$J         & 11203$\pm$300 & 4.0 (prefixed) &AT9\\
\hline\noalign{\smallskip}
\end{tabular}
\end{table}

\section{The resulting abundances}

The abundances used for computing the whole synthetic
spectrum of HR\,6000 from 3050\,\AA\ to 9460\,\AA\ 
are listed in column\,2 of Table\,3. The synthetic spectrum overimposed
on the observed spectrum is available at the website of
footnote\,1. The abundances  are based on
the ATLAS12 model with parameters 
\teff{}=12850\,K, \logg=4.1,
$\xi$=0.0\,km\,s$^{-1}$, and $v \sin i$=1.0\,km\,s$^{-1}$. 
They are shortly discussed below.

 The main results are 
the large underabundance  of  Si ([$-$2.9]) followed by that
of  C ([$-$1.8]). Other
underabundant elements are He ($\le$[$-$0.85]), N $\le$ ([$-$1.7]), 
O ([$-$0.5]), Ne([$-$0.9], Mg ([$-$1.2]), Al ([$-$1.7]),  S ([$-$1.5]),
Cl ($<$[$-$0.7]), Ca ([$-$0.2]), Sc ([$-$0.6]), V ([$-$1.1], 
Co $\le$ ([$-$0.8]), Ni ([$-$0.4]), and Sr ([$-$1.0]).  

On the other hand, the most overabundant element is Xe ([+4.5]), followed by
Hg ([+2.7]), and  P ([+2.3]:). Other
overabundant elements are Be ([+0.7]), Ti ([+0.25]), Mn ([+0.95]),
Fe ([+0.7]), Y ([+1.0]), and possibly Ba([+1.0]).  

\begin{table}
\caption[ ]{Final abundances $\log$(N$_{\rm elem}$/N$_{\rm tot}$) from the ATLAS12 model
with parameters \teff{}=12850\,K, \logg=4.1, $\xi$=0.0\,km\,s$^{-1}$.
Solar abundances are from Grevesse \& Sauval (1998).}
\centering
\begin{tabular}{lrrcrrcl}
\hline
\hline\noalign{\smallskip}
\multicolumn{1}{c}{Elem}&
\multicolumn{1}{c}{UVES}&
\multicolumn{1}{c}{IUE(old)}&
\multicolumn{1}{c}{IUE(new)}&
\multicolumn{1}{c}{Sun}&
\\
\hline\noalign{\smallskip}
\ion{He} & $-$1.90 & & &$-$1.05\\
\ion{Be} & $-$9.90 &$\le$$-$9.5&$-$9.90 &$-$10.64\\
\ion{B}  &         &$-$10.1 &$-$10.29&$-$9.49&\\
\ion{C}  & $-$5.32: &$\ge$$-$5.4&$-$5.82& $-$3.52\\
\ion{N}  &$\le$ $-$5.82 &$\ge$$-$6.5&$-$6.50& $-$4.12\\
\ion{O}  & $-$3.71 &$-$3.3& $-$3.70&$-$3.21\\
\ion{Ne} & $-$4.86 &&& $-$3.96\\
\ion{Na} & $-$5.71&&& $-$5.71\\
\ion{Mg} & $-$5.66 &$-$5.6&$-$5.66&$-$4.46\\
\ion{Al} & $-$7.30 &$-$7.7&$-$7.30&$-$5.57\\
\ion{Si} & $-$7.40 &$<$$-$5.7&$-$7.40&$-$4.49\\
\ion{P}  & $-$4.30 &$-$4.4$-$$-$5.0&$-$5.00&$-$6.59\\
\ion{S}  & $-$6.20 &$-$6.0$-$$-$5.5&$-$6.20&$-$4.71\\
\ion{Cl} & $\le$$-$7.24 &$-$4.3&$-$7.24& $-$6.54\\
\ion{Ca} & $-$5.88 &$-$5.2&$-$5.88&$-$5.68\\
\ion{Sc} & $-$9.50 &$-$11.0&$-$9.50&$-$8.87\\
\ion{Ti} & $-$6.77 &$>$$-$6.5&$-$6.77&$-$7.02\\
\ion{V}  & $-$9.14 &$-$6.7$-$$-$8.0&$-$9.14&$-$8.04\\
\ion{Cr} & $-$6.37 &$-$6.2&$-$6.37&$-$6.37 \\
\ion{Mn} & $-$5.60 &$-$5.3&$-$5.60&$-$6.65\\
\ion{Fe} &$-$3.85 &$-$3.8&$-$3.85& $-$4.54\\
\ion{Co} &$<$$-$7.92 &$\ge$$-$9.3&$-$7.92& $-$7.12\\ 
\ion{Ni} & $-$6.19&$-$7.0&$-$6.19&$-$5.79\\
\ion{Cu} &$\le$ $-$7.83&$\ge$$-$9.3&$-$7.83&$-$7.83\\
\ion{Zn} &        &$-$8.3&$-$7.44&$-$7.44\\
\ion{Ga} & $\le$$-$9.16&$-$8.4&$-$9.16&$-$9.16\\
\ion{Sr} &$-$10.07&&&$-$9.07 \\
\ion{Y}  &$-$8.80 &&&$-$9.80\\ 
\ion{Xe} & $-$5.40&&& $-$9.87\\
\ion{Ba} & $-$8.91?&&&$-$9.91\\
\ion{Hg} &$-$8.20 &$-$&$-$8.20&$-$10.91\\
\noalign{\smallskip}
\hline
\noalign{\smallskip}
\end{tabular}
\end{table}

Below we discuss the individual abundances:
 
{\it Helium}: From the analysis of the \ion{He}{i} lines 
4713\,\AA{} and 5875\,\AA{}, Catanzaro et al.\ (2004) derive an underabundance of
[$-$1.45], i.e. N(\ion{He})/N$_{tot}$=0.0032. We extended their analysis to
all the \ion{He}{i} lines previously analyzed 
in the HgMn star HD\,175640 by  Castelli \& Hubrig (2004b), except for 
\ion{He}{i} 4713\,\AA{} which is located in a gap of the observed spectral regions.
While all the \ion{He}{i} lines were fairly well predicted in HD\,175640 by the
same abundance, this is not a case in HR\,6000. In fact,
the abundance N(\ion{He})/N$_{tot}$ from the different lines 
ranges from 0.0125 to 0.002, i.e from [$-$0.80] to [$-$1.60].
The higher value reproduces the whole profile of \ion{He}{i} 3867.4\,\AA{} and  
only the wings of \ion{He}{i} lines at 4026.1\,\AA{} and 4387.9\,\AA{}, whose computed cores  
appear stronger than the observed ones (Fig.\,4).
 The same discrepancy  occurs also 
for the lines at $\lambda\lambda$ 4471.4, 4921.9, 5015.7\,\AA{}, whose
 wings 
are fitted by abundances equal to  0.005 and 0.006, while the cores would require 
a lower value.
The two weak lines at $\lambda\lambda$ 5875.6 and 6678\,\AA{} 
are rather well predicted for N(\ion{He})/N$_{tot}$ equal to 0.002 and 0.003, 
respectively. Finally,
the line at 7065.2\,\AA{} is partially blended both with telluric lines and with
\ion{Fe}{ii} 7065.163\,\AA{}. It can not be reproduced whichever is the abundance. 
The abundance from the remaining \ion{He}{i} lines, which are very weak,
depends on the assumed 
position for the continuum level. The comparison between observed
and computed  \ion{He}{i} profiles is available at the website of 
footnote 1.

Profiles of \ion{He}{i} like those observed in HR\,6000 were observed 
also in other
HgMn stars as for instance HR\,7361, HR\,7664 (Dworetsky, 2004) or Feige\,86 (Bonifacio
et al., 1995). This kind of anomalous profiles was interpreted as
evidence of helium stratification (Dworetsky, 2004; Bohlender, 2005).

No contribution of $^{3}$\ion{He}\,to $\lambda$ 6678\,\AA{} was observed.

\begin{figure}
\centering
\includegraphics[height=0.5\textwidth,angle=90]{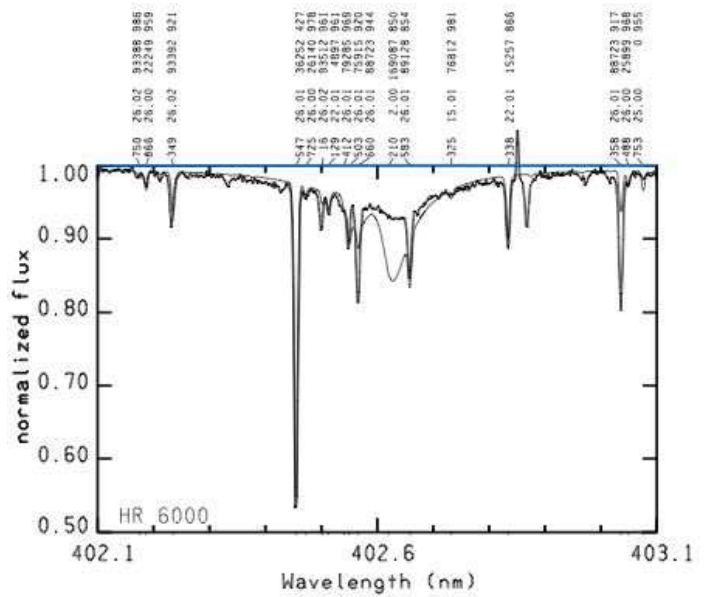}
\caption{Comparison of observed (thick line) and computed (thin line)
\ion{He}{i} 4026\,\AA{} profiles. The abundance
$\log$(N(He)/N$_{tot}$)=$-$1.90 which fits the wings  yields a 
too strong computed core.}
\label{fig4}
\end{figure}

{\it Beryllium}: The abundance was derived from the lines
at 3130.42 and 3131.06\,\AA{}, although the second one is heavy blended.

{\it CNO}: While Catanzaro et al.\ (2004) did not observe \ion{C}{ii} lines,
 we possibly identified very weak \ion{C}{ii} lines at 
$\lambda\lambda$ 3918.968, 4267.01, and 4267.26\,\AA{}.
 No nitrogen lines were observed. We deduced a
 carbon underabundance of [-1.80] and an  upper limit for nitrogen of 
[-3.40].  Oxygen is underabundant by [-0.50]. We examined
 all the \ion{O}{i} lines listed in Castelli \& Hubrig (2004b). 
The \ion{O}{i} infrared lines at $\lambda\lambda$ 7774,  8446, 
and 9260\,\AA{} are inadequately computed under the LTE assumption.

{\it Neon}: The abundance was fixed from \ion{Ne}{i} 7032.413\,\AA{}.

{\it Sodium}: There is a remarkable inconsistency between the abundance of
$-$3.31\,dex  from the two lines \ion{Na}{i} at 3302.368\,\AA{}
and 3302.978\,\AA{} 
and the almost solar abundance of $-$5.57\,dex of the \ion{Na}{i}  lines 
at $\lambda\lambda$ 5688.2, 8183.2 and 8194.8\,\AA{}. 
As already described by Catanzaro et al.\ (2004) the two stellar \ion{Na}{i} 
D lines  at  5889.951\,\AA{} and 5895.924\,\AA{} are completely hidden by
the interstellar lines. The most likely explanation is that the
lines at 3302\,\AA{} are also of interstellar origin. They have also been observed
in the spectrum of HD\,175640 as weak red-shifted absorptions with
the same wavelength displacement of +0.08\,\AA{} as the interstellar 
D-lines.

{\it Magnesium}: The estimated underabundance is [-1.20]. 
Magnesium abundance was mostly derived from \ion{Mg}{ii} 4481\,\AA{}. 
The \ion{Mg}{ii} doublet at 6545.942, 6545.994\,\AA{} is 
much broader than predicted and is also red-shifted. The most plausible
explanation is the lack of Stark broadening data in the computations.

{\it Aluminium}: The estimated underabundance is [-1.73] from 
the \ion{Al}{ii} lines at $\lambda\lambda$ 7042.06 and 7056.60\,\AA{}. 
No other lines, as \ion{Al}{ii}  at 
$\lambda\lambda$ 3944.009 and 3961.523\,\AA{}, were observed.

{\it Silicon}: The estimated underabundance is [-2.90]. 
Only weak  \ion{Si}{ii} lines  at $\lambda\lambda$
3853.665, 3856.018, and 3862.595\,\AA{} were observed.

{\it Phosphorus}: The overabundance of phosphorus can be inferred from most
of the \ion{P}{ii}  and \ion{P}{iii} lines, but  the real 
value is very difficult to be deduced
because the different lines give rather different abundances.
We measured the equivalent widths for a sample of 
22 \ion{P}{ii} and  4 \ion{P}{iii} lines. The average abundance 
from \ion{P}{ii} is $\log$(N(P)/N$_{tot}$)=$-$4.36$\pm$0.28 
for $\xi$=0.0~km\,s$^{-1}$, but 
the trend of the abundances with the equivalent widths is very large.
It disappears for $\xi$=6.0\,km\,s$^{-1}$.   The corresponding abundance
of $-$4.70$\pm$0.12 produces, for $\xi$=6.0\,km\,s$^{-1}$, extremely broad 
and shallow profiles completely different from the observe ones.

The abundance for $\xi$=0.0\,km\,s$^{-1}$ inferred from the \ion{P}{iii} 
lines is $\log$(N(P)/N$_{tot}$)=$-$4.57$\pm$0.27. 
Therefore, the \ion{P}{ii} abundance for $\xi$=0.0\,km\,s$^{-1}$ is 0.2\,dex larger 
than that obtained from \ion{P}{iii}, but lies withing the large
mean square error of the averages.    
The large uncertainty in the phosphorous abundance could be related
to NLTE effects which have been neglected here. Another possibility could be
the presence of phosphorus vertical abundance stratification.

The adopted \ion{P}{ii} and \ion{P}{iii} lines, together with the 
atomic data and the measured equivalent widths, are
listed on the website of footnote 1.

{\it Sulfur}: The sulfur underabundance of [$-$1.5] was derived from the same
lines used by Castelli \& Hubrig (2004b).

{\it Chlorine}: The absence of observed chlorine lines, like 
\ion{Cl}{ii} 4794.54\,\AA{}, implies an abundance less than $-$7.24~dex.

{\it Potassium:} Potassium  can be predicted by
adopting solar abundances. However, the slight blue-shift 
of $-$0.05\,\AA{} of the lines \ion{K}{i} $\lambda\lambda$ 7664.911, 7698.974\,\AA{},
the only ones observed in the spectrum,  may indicate 
an interstellar origin for them.

{\it Calcium}: All the \ion{Ca}{ii} lines listed in Castelli \& Hubrig (2004b) 
can be reproduced by an
underabundance of [-0.2], except for the lines of mult. 13  (8201.720\,\AA{} and 
8248.796\,\AA{}) which are predicted too strong. 
The blue wing of the K line at 3933.664\,\AA{} presents an additional
broad weak absorption. It is further discussed in Sect.\,6.3.
The lines of the \ion{Ca}{ii} infrared triplet are red-shifted by 0.14\,\AA{}.
They are discussed in more detail in Sect.\,6.1.

The \ion{Ca}{i} line at 4226.728\,\AA{} is very weak. It is
reproduced by solar abundance.  A possible interstellar
contribution can not be excluded.

{\it Scandium}: The  underabundance of [-0.6] was derived from the 
 \ion{Sc}{ii} lines listed in Castelli \& Hubrig (2004b). 

{\it Titanium}: The overabundance of [+0.25] was derived 
from the  \ion{Ti}{ii} lines listed in Castelli \& Hubrig (2004b).

{\it Vanadium}: The underabundance of [-1.1] was deduced from 
the lines of \ion{V}{ii} mult.~1 at 3100\,\AA{}.

{\it Chromium:} Most of the observed \ion{Cr}{ii} lines can be predicted by
solar abundance.   

{\it Manganese:} The abundance of $-$5.0\,dex  from the lines of \ion{Mn}{ii}
 mult.3 at 3440$-$3498\,\AA{} is larger by 0.6\,dex than that 
obtained from most of the \ion{Mn}{ii} lines 
lying longward of the Balmer discontinuity. This result indicates 
the possible occurrence of  vertical stratification for the abundance of this ion.
The abundance of $-$5.6\,dex  listed in Table\,3 is the one
obtained from the region longward of the Balmer
discontinuity.  Its value is the same as that presented by
Catanzaro et al.\ (2004).

{\it Iron}: The abundance was derived from the equivalent widths
of 11 \ion{Fe}{i} and 34 \ion{Fe}{ii} lines.
The lines together with the atomic data and the measured equivalent widths are
available on the website of footnote 1. 
The overabundance of [+0.7] gives rise in the spectrum to a very large 
number of \ion{Fe}{i}, \ion{Fe}{iii}, and especially  \ion{Fe}{ii} lines (see sub-section 6.4).

{\it Cobalt}: No \ion{Co}{ii} lines were observed in the spectrum, 
in particular
the line at 3501.717\,\AA{} is absent. Therefore only an upper limit of [$-$0.8]
was fixed.

{\it Nickel}: Most \ion{Ni}{ii} lines indicate an underabundance of [$-$0.4]. 

{\it Copper}: No lines of \ion{Cu}{ii} were observed,
in particular those at $\lambda\lambda$ 4909.734, 4917.892, and
4931.698\,\AA.  Because for 
solar abundance there are no predicted lines we assumed  solar abundances
as upper limit for this element. 

{\it Gallium}: No gallium lines were observed. If a feature
at $\lambda$ 5416.32\,\AA{} is identified as \ion{Ga}{ii} it 
would be reproduced by
$\log$(N(\ion{Ga})/N$_{tot}$)=$-$6.0. This abundance gives rise 
to other strong computed
\ion{Ga}{ii} lines with no observed counterparts.

{\it Strontium}: No \ion{Sr}{ii} lines were observed. Only a very
weak absorption is detectable at the position of
\ion{Sr}{ii} 4077.71\,\AA{}. It is reproduced by assuming 
an underabundance of [-1.0].

{\it Yttrium}: The overabundance of [+1.0] was derived mostly from 
the \ion{Y}{ii} line at 4900.12\,\AA{}.

{\it Zirconium}: No Zr lines were observed in the spectrum.
They are not predicted by assuming solar Zr abundance.

{\it Xenon}:  Xenon is the most overabundant element ([+4.5]).
The abundance was derived from the \ion{Xe}{ii} lines listed
in Castelli \& Hubrig (2004b) having $\log\,gf$ available from
the NIST database. They were taken from Wiese \& Martin (1980) (WM80).
 Furthermore,  we added in the line list
some lines with $\log\,gf$'s from Ryabchikova \& Smirnov (1989).
In addition, only for identification purposes,
all the lines with no $\log\,gf$ in the NIST database but with 
intensity $\ge$ 100 were included  in the line lists.
A guessed  $\log\,gf$ was assigned to them.

We noticed that the position of several \ion{Xe}{ii} 
lines is close to
that of unidentified stellar lines with the wavelength 
blue-shifted by 0.05 to 0.1~\AA\ from the
\ion{Xe}{ii} laboratory wavelength. Examples
are the lines at  
$\lambda\lambda$ 4180.10, 4208.48, 4209.47, 4238.25, 4245.38, 
4330.52, 4393.20, 4448.13, 4462.19, 5667.56\,\AA{}.
We have observed the same occurrence  also in the two  Xe rich HgMn stars  
Feige 86 and 46 Aql (Castelli \& Hubrig, in preparation).
If the \ion{Xe}{ii} identification is correct, the wavelength shift
could be due to some isotopic anomaly.
In fact, no isotopic composition was considered in our computations
owing to the lack of isotopic wavelengths for \ion{Xe}{ii}. There are
data only for $\lambda$ 6051.150\,\AA\ (Alvarez et al., 1979). 
For this line the isotopic wavelengths range from 6051.148\,\AA\ to 6051.152\,\AA.
Because the stellar line is  blue-shifted by 0.03\,\AA\ from the
laboratory wavelength 6051.150\,\AA\ (Hansen \& Persson, 1987)
some isotopic anomaly can not be invoked to explain
the observed shift.

{\it Barium}: We were not able to draw any conclusive results for barium.
There is a weak absorption at 4933.93\,\AA{}, but if we identify it
as \ion{Ba}{ii} 4934.076\,\AA{}, this barium line would be
blue-shifted by 0.15\,\AA{}.
The wavelength in the spectrum would correspond to the Ritz wavelength
4933.970\,\AA{} rather than to the observed wavelength 4934.077\,\AA{}, as they 
are  listed in the NIST database.  
Any isotopic anomaly has to be excluded because the shortest 
isotopic wavelength 
is that of the hyperfine component of  $^{137}$Ba at 4934.054\,\AA{}
(McWilliam, 1998).

{\it Mercury}: Only \ion{Hg}{ii} at 3984\,\AA{} is observable. 
It can be predicted
by assuming an overabundance of [+2.7] and
a non-terrestrial isotopic composition.  The Hg isotopic anomaly is
discussed in Sect.~6.1.

\subsection{Comparison with the abundances from the IUE spectra}

Column 3 of Table\,3 lists the abundances that were derived from IUE spectra 
by Castelli et al.\ (1985). The analyzed images were SWP04588 
 and LWR03979.
We revised these abundances by using the same ATLAS12 model adopted
for the analysis of the UVES spectra and the same images that we 
downloaded from the MAST 
archive\footnote{http://archive.stsci.edu/iue/}.
 Also updated line lists, as compared
with those used by Castelli et al.\ (1985), were used. Column\,4 of Table\,3
lists the revised abundances from the ultraviolet.

The IUE spectra are very noisy and the resolving power is too low 
(R=$\sim$12000-13000) to allow one to perform
an accurate analysis. Almost all very numerous lines are
blended either with known or with unknown components.
In addition,
the never adequately corrected distortions of the echelle orders make the drawing
of the continuum a very problematic task. The large uncertainty in
the continuum position  affects both the shape and the
strength of the spectral lines.
 For all these reasons 
the abundances that we derived from the IUE spectra have always been  
considered as estimates rather than definitive values. 

Table\,3 shows  that the difference IUE(new)$-$IUE(old)
between  revised  and  Castelli et al.\ (1985) abundances
 is larger than 0.4\,dex for
\ion{C}, \ion{O}, \ion{Cl}, \ion{Ca}, \ion{Sc}  \ion{V}, \ion{Ni}, 
\ion{Zn}, and \ion{Ga}.
The abundances from UVES spectra reproduce the IUE 
spectrum (within all  the above outlined limits)  for all the elements 
except for carbon and phosphorous, for which the differences
UVES-IUE(new) are $+$0.5\,dex and +0.7\,dex
respectively.   
While the analysis of the carbon abundance from the visible spectral region is based only on
very weak blended lines and therefore is rather uncertain,
the phosphorous discrepancy confirms the difficulty
in fixing the phosphorus abundance from the different lines.

The synthetic spectrum overimposed on the IUE spectra is available
at the website of footnote\,1. It was computed with the
abundances listed in column\,4 of Table\,3.

\section{Line spectrum peculiarities}

In the following sub-sections we discuss isotopic anomalies observed for Hg 
and Ca and the emissions observed for a few lines of  \ion{Mn}{ii}, 
\ion{Cr}{ii}, and \ion{Fe}{ii}.
 We also discuss the huge quantity of lines which could not be identified.

\subsection{Isotopic anomalies}

Isotopic anomalies in HR\,6000 were observed
for the \ion{Ca}{ii} infrared triplet and for \ion{Hg}{ii} at 3984\,\AA{}.
No  $^{3}$\ion{He}/$^{4}$He isotopic anomaly was detected 
in the UVES spectra.

{\it \ion{Ca}{ii} infrared triplet}: The lines of the infrared 
triplet at $\lambda\lambda$\,8498.023, 8542.091,  and 8662.141\,\AA{} are 
red-shifted by 0.14\,\AA{} and appear broader and stronger than the predicted 
ones. The wavelength red-shift of the infrared triplet observed in several
CP stars (Cowley et al., 2007) was interpreted
by Castelli \& Hubrig (2004a) as due to a \ion{Ca}{} isotopic composition  different
from the terrestrial one.

{\it \ion{Hg}{ii} 3984}:
 \ion{Hg}{ii} at 3984\,\AA{}, although weak, was detected in the UVES spectra.
The isotopic composition that best reproduces the observed line 
for $\log$(N(\ion{Hg})/N$_{tot}$)=$-$8.20 is 
that listed in column\,3 of Table\,4. 
For comparison,  the terrestrial isotopic composition 
(Anders \& Grevesse 1989)  is given in column\,5 of
Table\,4. In HR\,6000 the isotope 204 is the most abundant one.
The profiles computed with the stellar and terrestrial isotopic compositions
are compared with the observed profile in Fig.\,5. 
 We note that the \ion{Hg}{ii} line is blended with
\ion{Fe}{i} 3983.956\,\AA{}.

\ion{Hg}{ii} at 1942.3\,\AA{} can be observed in the IUE spectra,
but the isotopic shifts are too small to enable us to confirm the
isotopic composition deduced from the line at 3984\,\AA{}.

\begin{table}
\caption{Mercury isotopic (iso) and hyperfine (hfs) composition in \% 
for \ion{Hg}{ii} at 3984\,\AA{}.}
\centering
\begin{tabular}{lr|rr|rrlllllll}
\hline
\hline\noalign{\smallskip}
Isotope     &
\multicolumn{1}{c}{$\lambda$}& 
stellar & $\log\,gf$ & terr.&$\log\,gf$& hfs.\\
            &         &\multicolumn{1}{c}{\%} &
\multicolumn{1}{c}{iso}& 
\multicolumn{1}{c}{\%}&
\multicolumn{1}{c}{iso}& 
\multicolumn{1}{c}{\%}& \\
\hline\noalign{\smallskip}
196&3983.769         &0.15    & -2.81            & 0.15 &$-$2.81\\
198&3983.838         &0.77    &-2.11             & 9.97 &$-$1.00\\
199a&3983.838        &20.00   &-0.70             &16.87 &$-$0.77&7.14\\
199b&3983.849        &20.00   &-0.70             &16.87 &$-$0.77&9.71\\
200&3983.909         &6.87    &-1.16             &23.10 &$-$0.64\\
201a&3983.930        &5.00    &-1.30             &13.18 &$-$1.32&4.80\\    
201b&3983.941        &5.00    &-1.30             &13.18 &$-$1.08&8.30 \\
202&3983.990         &10.00   &-1.00             &29.86 &$-$0.52\\
204&3984.071         &57.20    &-0.24            &6.87  &$-$1.16\\
\hline
\noalign{\smallskip}
\end{tabular}
\end{table}

\begin{figure}
\centering
\includegraphics[height=0.5\textwidth,angle=90]{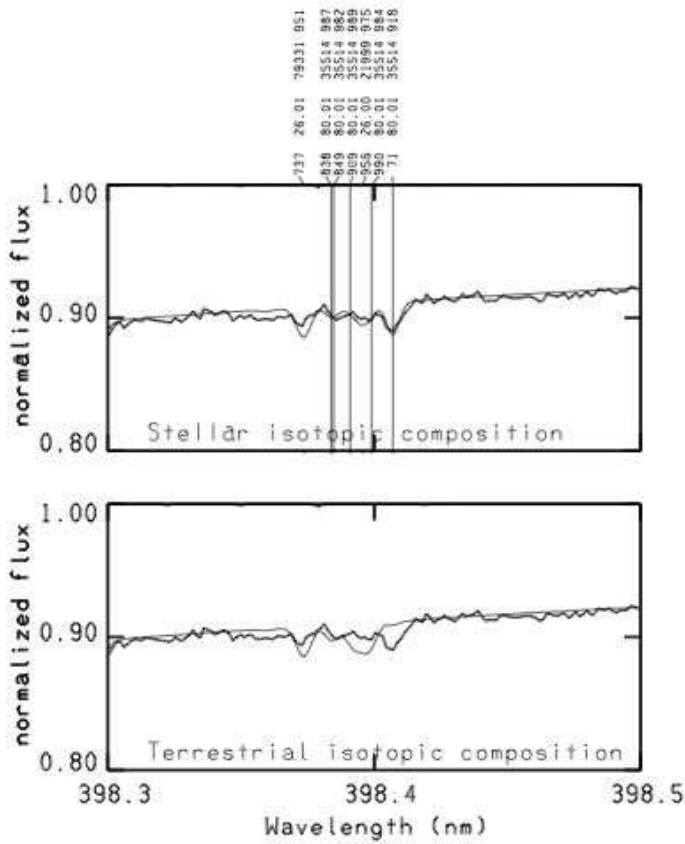}
\caption{The observed profile (thick line) of Hg~II 3984~\AA{} is compared 
in the lower panel with the
profile computed with a terrestrial isotopic composition (thin line) and
in the upper panel with the profile computed with the isotopic mixture listed in column\,3 of table\,4
(thin line). }
\label{fig5}
\end{figure}

\subsection{Emission lines}

Emission lines were observed for \ion{Mn}{ii}, \ion{Cr}{ii}, and \ion{Fe}{ii}.
 We assumed that the studied lines are true emissions only if they were 
observed in both
UVES spectra taken at two different epochs. Emission lines are listed in 
Table\,5.

{\it Emission lines of Mn~II}: Some lines of \ion{Mn}{ii} multiplet 13 have 
been observed either in 
emission ($\lambda\lambda$ 6125.863, 6126.218, 6126.510, 6131.923\,\AA{}) 
or much weaker than predicted ($\lambda\lambda$ 6122.434, 6122.810, 
6128.734\,\AA{}). Some others lines
($\lambda\lambda$ 6129.033, 6130.796, 6131.016\,\AA{})
are not observed in the spectrum taken on
September 2005, while they appear as weak emissions in the spectrum  
of March 2006. This fact may suggests the presence of some variability 
in the emissions. Finally, the two very weak \ion{Mn}{ii} lines of mult. 13 
at 6123.160\,\AA{} and 6129.254\,\AA{} are in absorption and agree rather well 
with the predicted lines.

Other weak \ion{Mn}{ii} emissions occur  at  6446.337\,\AA{} (\ion{Mn}{ii}, 
mult.\,19) which is blended with \ion{Fe}{ii} 6446.410\,\AA{}, and at
7219.968, 8495.229, 8565.819, 8769.175\,\AA{}.  An emission could also
affect the line at  8695.208\,\AA, which is observed much weaker than 
predicted, and the line at 8819.584\,\AA{} which, although predicted,
is not observed. We note that 
there is a rather good agreement for this last line between 
the observed and computed profiles in the HgMn star HD\,175640 
(Castelli \& Hubrig, 2004b), even considering
that no hyperfine structure is available for computations.  

{\it Emission lines of \ion{Cr}{ii}}:
A few \ion{Cr}{ii} lines arising from  high excitation levels and having large 
$\log\,gf$'s appear clearly in emission. 
These lines were also found in emission in 
HD\,175640 (Castelli \& Hubrig, 2004b). 

{\it Emission lines of \ion{Fe}{ii}}:
There are no clear emissions for \ion{Fe}{ii}. 
The most probable ones are listed
in Table\,5.  However, there is a set of observed
lines much weaker than predicted. This disagreement could be explained either due to 
incorrect $\log\,gf$\,s' 
or to emissions filling the absorptions. 
They occur at $\lambda\lambda$ 6145.611, 6214.948,
6328.498, 6340.841, 6349.601, 6419.626, 6436.598, 6451.094, 6479.607, 6480.807,
6518.774, 6524.726, 6576.153, 6576.170\,\AA{}.

The phenomenon of weak emission lines at optical wavelengths for main-sequence
B-type stars has been noted from observations of relatively few stars.
Presently, explanations of this phenomenon have been put forward in the
context of the non-LTE line formation (Sigut, 2001) and possible
fluorescence mechanisms (Wahlgren \& Hubrig, 2000).

\begin{table*}
\caption[ ]{Emission lines observed in HR\,6000.} 
\centering
\begin{tabular}{llrrllllll}
\hline
\hline
\noalign{\smallskip}
\multicolumn{1}{c}{$\lambda$(\AA{})}&
\multicolumn{1}{c}{Element}&
\multicolumn{1}{c}{Mult.}&
\multicolumn{1}{c}{$\log~gf$}&
\multicolumn{1}{c}{$\chi_{low}$}&
\multicolumn{1}{l}{Notes}
\\
\hline
\noalign{\smallskip}
5518.08    & ?& \\
6112.906  & \ion{Fe}{ii}? &$-$&$-$3.753&90300.625& not predicted.\\
6122.434  &\ion{Mn}{ii}& 13 & 0.950 & 82136.400 & absorption filled by emission?\\
6122.810  &\ion{Mn}{ii}& 13 & 0.084 & 82136.400 &absorption filled by emission?\\
6125.863  &\ion{Mn}{ii}& 13 & 0.783 & 82144.480 \\
6126.218  &\ion{Mn}{ii}& 13 & 0.230 & 82144.480 \\
6126.510   &\ion{Mn}{ii}& 13 &$-$0.791&82144.480\\
6128.734  &\ion{Mn}{ii}& 13 & 0.588 & 82151.160 &absorption filled by emission?\\
6129.033   &\ion{Mn}{ii}& 13 & 0.208 & 82151.160 & variable?\\
6130.796   &\ion{Mn}{ii}& 13 & 0.354 & 82155.840 & variable?\\
6131.016   &\ion{Mn}{ii}& 13 & 0.053 & 82155.840 & blend with \ion{Fe}{ii} in absorption\\
6131.923   &\ion{Mn}{ii}& 13 & 0.053 & 82158.170 &\\
6158.621   &\ion{Cr}{ii}& $-$& 0.718 & 89174.080 &\\
6182.340   &\ion{Cr}{ii}& $-$& 0.452 & 89336.890 &\\
6446.337 &\ion{Mn}{ii}& 19 &$-$0.663& 98423.300& on the blue wing of \ion{Fe}{ii} 6446.410\,\AA{}\\
6526.302& \ion{Cr}{ii} &$-$ & 0.173 & 89885.080\\
6585.241& \ion{Cr}{ii} &$-$ & 0.829& 90850.960& red-shifted\\
7174.368& \ion{Mn}{ii} &$-$ & 0.397& 85960.460\\
7175.815& \ion{Mn}{ii} &$-$ &$-$0.023   & 85960.460\\
7219.968& \ion{Mn}{ii} & $-$ &0.768 &86057.440& \\
7222.639& \ion{Mn}{ii} & $-$ &0.252 &86057.440&\\
8495.229& \ion{Mn}{ii} & $-$ &0.198 &85960.460&\\
8565.819& \ion{Mn}{ii} & $-$ &0.278 &86057.440& \\
8695.208 & \ion{Mn}{ii} &$-$&0.577 &74560.010&absorption filled by emission?\\
8740.349 & \ion{Fe}{ii} &$-$&0.002 &99068.450\\
8758.410 &  ?&\\
8769.175& \ion{Mn}{ii}&$-$ & 0.491 &74560.010& \\
8819.584 & \ion{Mn}{ii}&13& 0.347 &74560.010&  absorption filled by emission?\\
8839.005& \ion{Fe}{ii}&$-$&&&  J78, unclassified$^{a}$\\
9020.127& \ion{Fe}{ii}&$-$&&&  J78, unclassified$^{a}$ \\
9085.11&?&\\
9208.4&? &\\
\noalign{\smallskip}
\hline
\noalign{\smallskip}
\multicolumn{2}{l}{$^{a}$J78=Johansson(1978)}\\
\end{tabular}
\end{table*}

\subsection{Unidentified lines}

There is a huge number of unidentified lines in the spectrum of HR\,6000.
Their list is available at the web-address given in footnote\,1.
For instance, most impressive regions are $\lambda\lambda$ 4404-4411\,\AA{} 
and 5100-5300\,\AA{},
which are overcrowded with unidentified absorptions. We estimate that most of them
are due to \ion{Fe}{ii}. In fact, a comparison with the spectrum of 
HD\,175640 (Castelli \& Hubrig, 2004b) has shown unidentified lines at 
the same wavelengths also in this star, although less intense 
than in HR\,6000, in accordance with
the different iron abundances (Fig.\,6). 
Therefore the unidentified lines are not a special peculiar characteristic of
HR\,6000. 

In the Kurucz line list for \ion{Fe}{ii}
(Kurucz, 2005)\footnote{http://kurucz.harvard.edu/atoms/2601/gf2601.pos}
there are numerous transitions 
between observed levels with wavelengths close to those of the unidentified lines, 
but with so low $\log\,gf$\,'s
that none corresponding absorption is predicted. The comparison of $\log\,gf$\,'s
from Kurucz  with the $\log\,gf$\,'s  from 
Raassen \& Uylings (1998) has shown agreement in the two
determinations. Therefore, either both determinations are off by several
orders of magnitudes or other transitions at the same wavelength  occur.
This possibility seems partly justified when the  Kurucz  
line list including lines from both observed and predicted 
levels is used\footnote{http://kurucz.harvard.edu/atoms/gf2601/gf2601.lines0600}.
In fact, numerous  \ion{Fe}{ii} lines due to transitions between predicted
levels have $\log\,gf$\,'s large enough to produce observable absorptions.
Unfortunately, the 
uncertainty in their wavelength may be as large as 1\,\AA{} and
also the corresponding $\log\,gf$'s may be affected by rather large
errors. 
Table\,6 is an example of tentative line identification for the 
unidentified lines in the interval 5130-5136\,\AA{}. 
As it is shown in Fig.\,6 and in the lower panel of Fig.\,7, out of displayed 17 spectral lines
only three lines can be identified as arising 
from observed levels. They are  \ion{Fe}{ii} 5131.210\,\AA{}, \ion{Fe}{ii} 5132.669\,\AA{}, and
\ion{Fe}{i} 5133.688\,\AA{}. The middle panel of  Fig.\,7 shows that
 more lines would be identified if several 
$\log\,gf$\,'s listed in  col.\,5 of Table\,6   are  
replaced by those given in col.\,7. They were  
derived by fitting the computed lines to
the observed lines for the stellar iron abundance 
of $-$3.85\,dex.
Upper panel of Fig.\,7 shows that the number of identified lines 
increases  if the \ion{Fe}{ii} lines from predicted levels 
listed in cols.\,8 and 9 of Table\,6 are added.
The intensity of the lines from predicted levels is 
usually lower than that of the observed lines.

 There is a  
remarkable coincidence between several  \ion{Fe}{II} lines measured 
by Johansson (1978) (Table\,6, col.\,3) and the unidentified lines in our spectra.
However, they are not predicted by the synthetic spectrum either because they are missing in the line lists or owing to
their too low assigned $\log\,gf$.

\begin{figure*}
\centering
\includegraphics[height=\textwidth,angle=90]{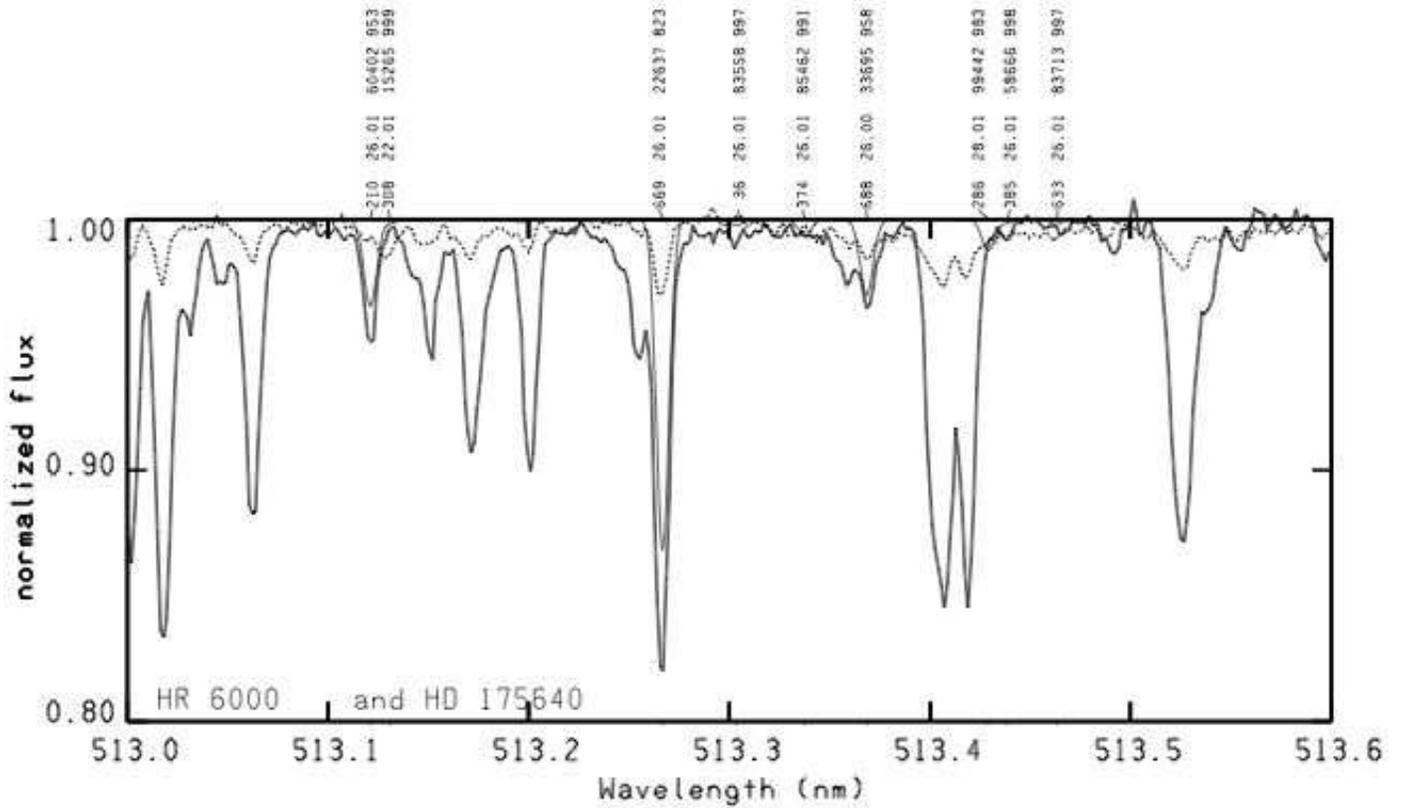}
\vskip -0.5 cm
\caption{The spectrum of HR\,6000 (full thick line) is compared with that of HD\,175640 
(dashed line) in a
region crowded with lines not predicted by the synthetic spectrum (thin line). The labels
indicate the only identifications in this spectral range.} 
\label{fig6}
\end{figure*}

\begin{table*}
\caption[ ]{Unidentified lines in the region 5130-5136\,\AA{}. } 
\centering
\begin{tabular}{lllrllrlrll}
\hline
\hline
\noalign{\smallskip}
\multicolumn{1}{c}{$\lambda$(obs)}&
\multicolumn{1}{c}{Identif.}&
\multicolumn{1}{c}{$\lambda$}&
\multicolumn{1}{c}{$\lambda$}&
\multicolumn{1}{c}{$\log\,gf$}&
\multicolumn{1}{c}{$\log\,gf$}&
\multicolumn{1}{c}{$\log\,gf$}&
\multicolumn{1}{c}{$\lambda$}&
\multicolumn{1}{c}{$\log\,gf$}&
\\
\multicolumn{1}{c}{(\AA{})}  &   & 
\multicolumn{1}{c}{J78$^{a}$}&
\multicolumn{1}{c}{K03$^{b}$}&
\multicolumn{1}{c}{K03$^{b}$}&
\multicolumn{1}{c}{RU$^{c}$} &Modified&
\multicolumn{2}{c} {predicted(K03)$^{b}$}&\\
\hline
\noalign{\smallskip}
5130.0  & \ion{Fe}{ii} & 5130.011  & 5130.013&   $-$5.590& $-$5.358 &$-$1.790& 5130.006& 1.172\\  
5130.18 & \ion{Fe}{ii} & 5130.177  &  missing&&&&5130.196 &$-$0.673\\     
        & \ion{Fe}{ii} &           &          &&&&5130.198 &$-$0.573\\ 
5130.3  & \ion{Fe}{ii} &           &5130.291    &  $-$6.352 &         &$-$1.252 \\                     
5130.45 & \ion{Fe}{ii} &              &             &         &       & &5130.427&$-$0.568\\
5130.60 & \ion{Fe}{ii} & 5130.605 uncl.  & 5130.618  & $-$6.012& $-$5.562 &$-$0.312&5130.542 &$-$0.120\\  
5131.4  & \ion{Fe}{ii} &              &         &          &       & &5131.478 &$-$0.570\\
5131.5  & \ion{Fe}{ii}  &            &         &          &        &&5131.650 &$-$0.508\\
5131.7  & \ion{Fe}{ii} &            &5131.706 &   $-$5.127&           &$-$0.127&5131.878 &$-$0.727\\  
        & \ion{Fe}{ii} &            &  5131.788 &   $-$5.023&           &$-$0.723&\\ 
5132.0  & \ion{Fe}{ii} &            &  5132.002 &   $-$4.648& $-$4.590  &$-$1.648&5131.953&$-$0.222\\ 
5132.55 & \ion{Fe}{ii} &                &           &           &  &      &5132.554&$-$0.190\\  
5133.59 & \ion{Fe}{ii} &            &5133.589 &   $-$3.961&             &$-$1.061&5133.475&0.220\\ 
5134.07 & \ion{Fe}{ii} &            &5134.082 &   $-$5.805&            & $-$0.105&&\\ 
5134.2  &  ?    &\\
5134.9  &  ?    &\\
5135.25 &\ion{Fe}{ii}  & 5135.268   & missing &           &      &            &\\
        &\ion{Fe}{ii}  &            &5135.187 & $-$6.671 &           &$-$2.671\\ 
        &\ion{Fe}{ii}  &            &5135.297 & $-$5.119& $-$4.394   &$-$1.719\\
5135.35 & ?     &\\
5135.95 & ?     & \\
\hline
\noalign{\smallskip}
\multicolumn{8}{l}{$^{a}$J78=Johansson (1978)}\\
\multicolumn{8}{l}{$^{b}$K03=Kurucz (2003)}\\
\multicolumn{8}{l}{$^{c}$RU=Raassen \& Uylings (1998)}\\ 
\end{tabular}
\end{table*}

\begin{figure*}
\centering
\includegraphics[height=\textwidth,angle=90]{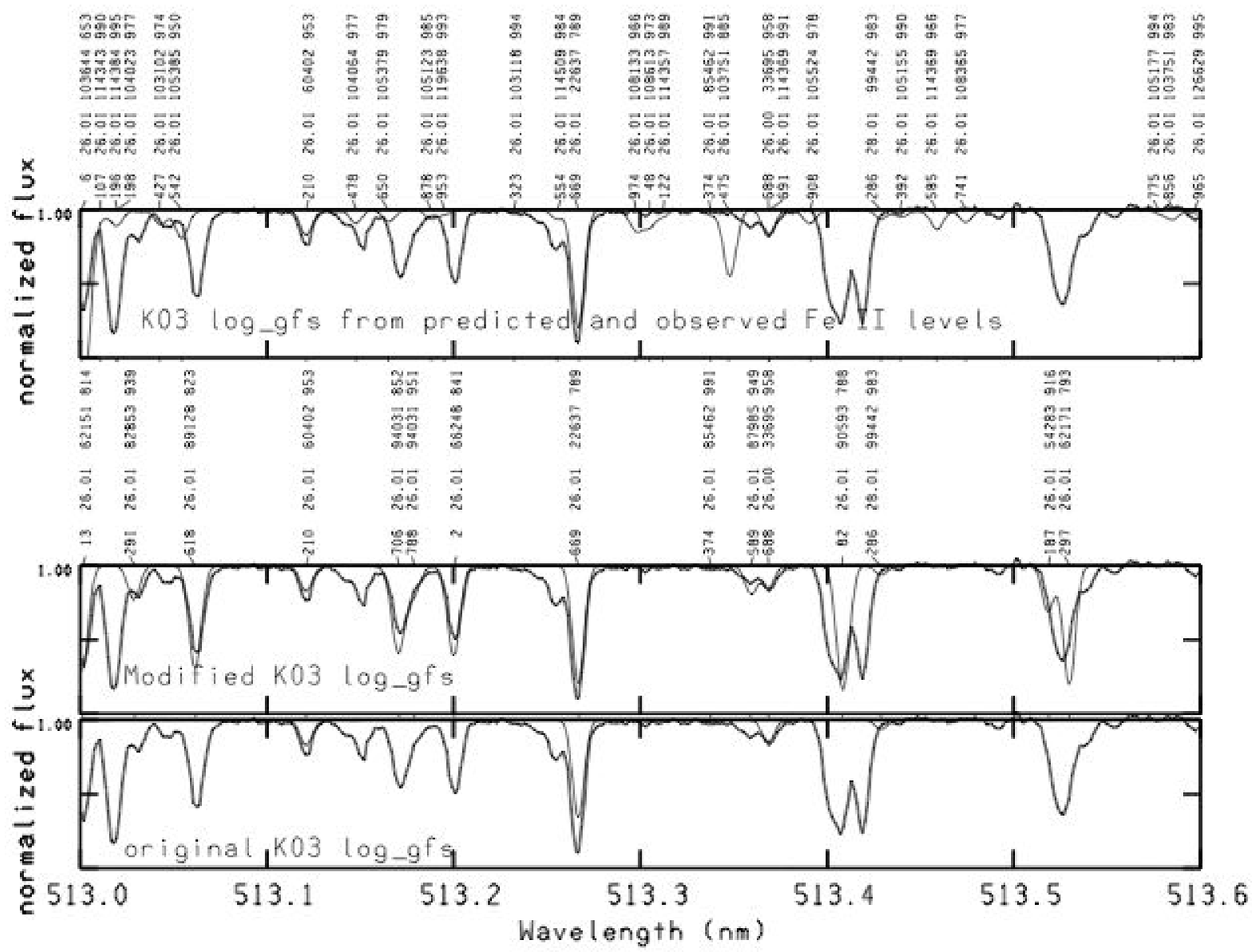}
\caption{The observed spectrum of HR\,6000 (thick line) is compared with synthetic spectra
(thin line) computed with the different choices of \ion{Fe}{ii} $\log\,gf$'s.
The lower panel shows the synthetic spectrum as predicted by the Kurucz (2003) line lists
when only transitions from observed levels are considered. 
Middle panel shows the synthetic spectrum when modified $\log\,gf$'s for several \ion{Fe}{ii} lines are adopted
(Table\,6, col.\,7). Upper panel  displays the synthetic spectrum computed by considering \ion{Fe}{ii}
transitions from both observed and predicted levels.}
\label{fig7}
\end{figure*}

\section{The T\,Tauri companion}

Van den Ancker et al.\ (1996) suggested the presence of a T\,Tauri companion
for HR\,6000 in order to explain the strong X-ray emission observed by
Zinnecker \& Preibisch (1994) and
the infrared excess that they inferred
from the comparison of the observed energy distribution with that computed 
for the best fitting model having
parameters \teff{}=14000\,K, \logg=4.3.
They assumed a temperature of 3500\,K for the T\,Tauri central star, an 
accretion rate of 2$\times$10$^{-7}$ M$_{\odot}$\,yr$^{-1}$,
and a V magnitude of about 13.5, so that the companion would be about 
6\,mag fainter than HR\,6000 and can not contribute significantly to 
the observed visual spectrum.

There are no clear spectroscopic traces of a second star in the
spectrum of HR\,6000, except for a broad weak absorption  at $\lambda\lambda$\,6707-6708\,\AA{}
which, although of different intensity in the two spectra observed
on September 2005 and March 2006, exhibits a very similar shape.
 The synthetic spectrum of HR\,6000 considered as a single star 
is flat in this region except for two weak \ion{Fe}{ii} lines at   
6707.232\,\AA{} and 6707.539\,\AA{} which do not cover the whole
observed absorption (Fig.\,8, lower panel).
We could argue that the feature is \ion{Li}{i} 6707.7 which
is a key element for the classification of T Tauri stars
(Krautter et al., 1997). 
If such a star is present, it should be 
similar to a weak emission line T\,Tauri star (WTT). WTT stars 
are defined as low-luminosity T\,Tauri stars with strong
X-ray emissions and with an optical counterpart showing pre-main-sequence 
characteristics, in particular  strong \ion{Li}{i} absorption at 6707\,\AA{}
and very weak or also absent emissions in H$\alpha$. Furthermore, the 
radial velocity of the WWT stars must be consistent with that of the 
members of the molecular cloud to which they belong (Bertout, 1989).

\begin{figure}
\centering
\includegraphics[height=0.5\textwidth,angle=90]{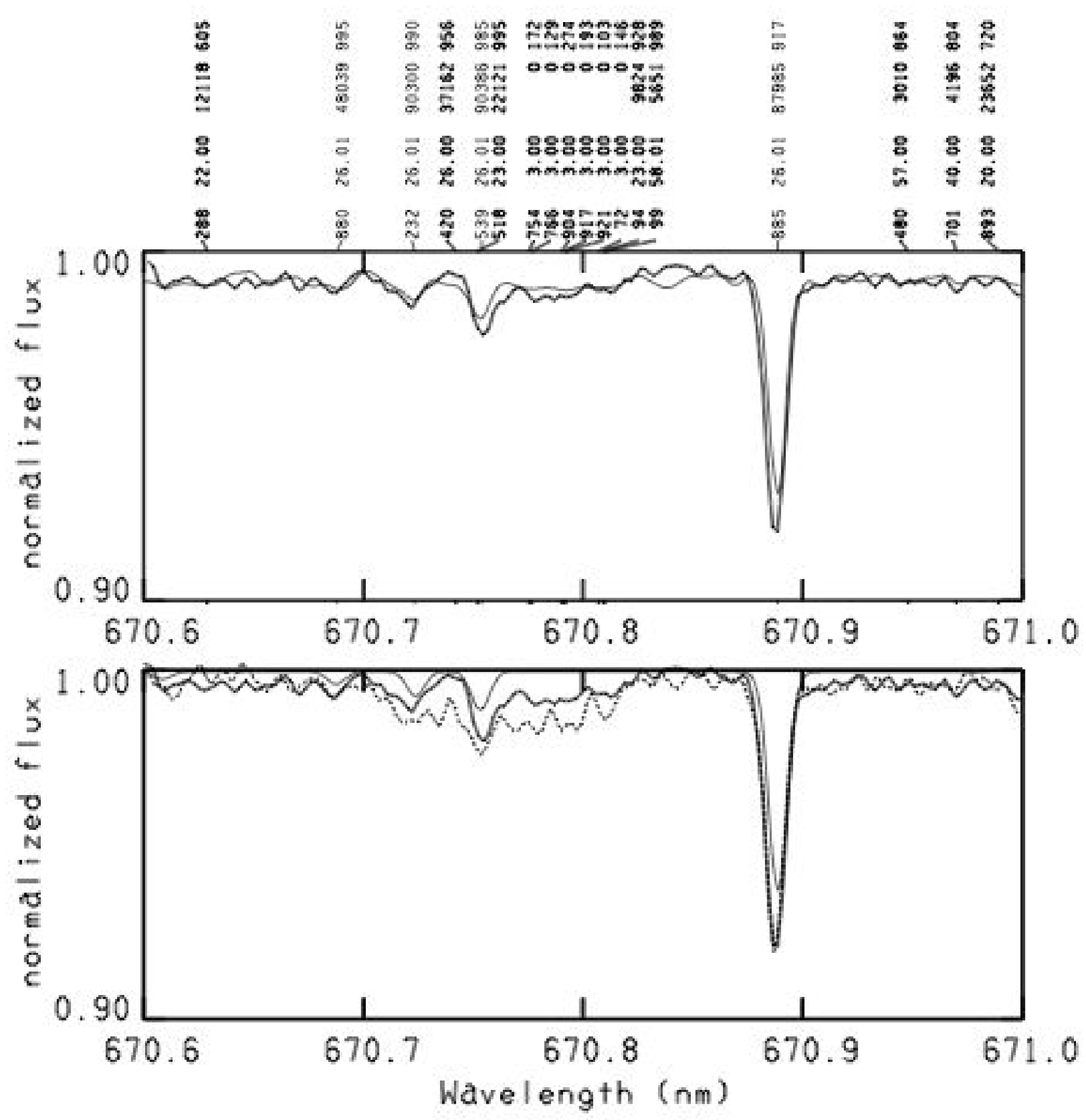}
\caption{Observed and computed profiles in the \ion{Li}{i} region. Lower panel 
shows the spectra observed on September 2005 (thick line) and on March 2006
(dashed line) overplotted on the synthetic spectrum of 
HR\,6000 considered as a single star (thin line). Upper panel compares the spectrum
observed on September 2005 (thick line) with the spectrum computed by
combining the synthetic spectrum of HR\,6000  with a spectrum computed
for \teff=3500\,K, \logg=4.0, [M/H]=0.0.  The thin  labels 
indicate atomic lines of HR\,6000 and the thick labels those of the other star. 
No labels for molecular
lines are printed, although molecular lines were considered for computing 
the synthetic spectrum of the cool star.
  }
\label{fig8}
\end{figure}

Upper panel of Fig.\,8 shows that the observed
spectrum is much better reproduced by a convolution of the
synthetic spectrum of HR\,6000 with a synthetic spectrum
computed for \teff{}=3500\,K, \logg=4.0 and solar abundances.
The computed spectrum of the cool star was not shifted in radial velocity 
relatively to that of HR\,6000.
In this case the contribution of \ion{Li}{i} 6707.7 
improves the agreement between the computed spectrum
and the observed spectrum whose continuum was lowered by 1.25\%.
We note that the main contribution 
to the whole absorption  comes from the TiO lines whose labels 
are not plotted in Fig.\,8, due to their huge number.
All the considerations relative to Fig.\,8 can be extended to 
the whole spectrum, starting from about 5800\,\AA{}.

To model the combined spectrum shown in the upper panel of Fig.\,8 
we assumed a luminosity ratio 
L$_{\star}$/L$_{T Tauri}$=220, which is among the
values suggested by van den Ancker et al.\ (1996).
 When the distance d=140$\pm$20\,pc adopted for HR\,6000 by van den Ancker  
et al.\ (1996) is replaced by the
Hipparcos distance d=240$\pm$48\,pc, the stellar luminosity computed with
formula (1) from van den Ancker et al.\ (1996) is L$_{\star}$=324L$_{\odot}$.
In this case the luminosity
of the T Tauri star becomes L$_{T Tauri}$=1.5L$_{\odot}$.
For the computation of the binary spectrum the ratio of the stellar 
radii (R$_{TTau}$/R$_{HR6000)}$)$^{2}$
was set to be equal to 0.8 in accordance with the Stefan-Boltzmann law.
A change of the luminosity ratio affects only the level of the computed 
continuum. 
Several experiments 
made with other choices of the parameters with \teff{} ranging 
from 3500\,K to 4500\,K and \logg\ ranging from 4.0 to 1.5 
have shown that the above results are not significantly modified.

In any case, although we can not exclude in an absolute way that
the spectrum of HR\,6000 is contaminated by that of a WTT star, 
we can not exclude as well
that we simply observe  spectral noise rather than weak features
of a secondary spectrum. In fact, the inspection of UVES spectra
of other CP stars observed during the same runs has shown 
the presence of weak features very similar to those we just
discussed here for HR\,6000.

\section{Conclusions}

The present study of HR\,6000 has led to the following results: 
the measured radial velocity indicates that 
HR\,6000 most likely belongs to the Lupus cloud so that its
age is of the order of 10$^{7}$ years as it was found  
for the Lupus\,3 cloud (James et al., 2006).
Interstellar lines from the cloud can be observed in
the spectrum from the ultraviolet to the visible, with 
displacements of the order of 
$-$0.05 $\div$ $-$0.15\,\AA{} from the stellar lines.

UVES spectra observed with a six months time interval do not show
spectrum variability for both stellar and interstellar 
lines, but a low variability of 
1.2\,km\,s$^{-1}$ in the radial velocity  and marginal variabilities 
of the weak spectral features which are however at the level of the noise.

Different methods for the parameter determination 
have led to large differences in \teff{}.  While the  
gravity values are found in the range between 4.1 and 4.4\,dex
with errors of the order of 0.1$\div$0.2\,dex, 
\teff{} changes from
13900$\pm$300\,K, (if obtained from the UBV$\beta$ and uvby$\beta$ photometry), 
down to 12850$\pm$50\,K  when determined from the Balmer profiles, 
and can become as low as 11200$\pm$300\,K when derived from the  VI$_{c}$J photometry.
Furthermore, as far as the Balmer profiles are concerned, the parameters which
well reproduce H$_{\beta}$, H$_{\gamma}$, and H$_{\delta}$
predict  too narrow H$_{\alpha}$ wings.
All these discrepancies in the parameter determinations
could be accounted for by the presence of the Lupus cloud with 
could affect the colors in such a way to invalidate
the standard reddening relations.
Also a spectral contamination by a weak-emission T\,Tauri (WTT)
companion, as suggested by van den Ancker et al.\ (1996),
can not be excluded in an absolute way,
although the only spectroscopic sign of its possible
presence is a weak variable broad absorption at the
position of \ion{Li}{i} 6707\,\AA\ and a somewhat better
agreement between  observed and computed spectra  
longward of 6000\,\AA{} for a combined synthetic spectrum  
obtained from that  HR\,6000  and one
computed for \teff{}=3500\,K, \logg=4.0, [M/H]=0.0. 
This companion could be either physically related with HR\,6000 
to form a close binary system or could be located in the foreground
of HR\,6000. 
Finally, the different parameters derived from the different determinations
could be due to  the extremely peculiar nature of HR\,6000 which may
make the classical model atmospheres inadequate to reproduce all the 
observed stellar quantities.
In fact, the abundance stratification inferred for
some elements, helium in particular, would require more refined models 
in which the hypothesis of constant abundances throughout the atmosphere
is dropped.

By using an ATLAS12 model with parameters \teff=12850\,K, \logg=4.1,
$\xi$=0.0\,km\,s$^{-1}$ we have computed a synthetic spectrum for 
HR\,6000 from 3050\,\AA\ to 9460\,\AA\ and have compared it with
the observed spectrum. This comparison has shown that
most of the elements lighter than Ca are significantly
underabundant, except for Be, Na, and P.
The Si underabundance ([-2.9]) is remarkable because it is
even lower than that of 46 Aql ([$-$1.0]) which was claimed 
by Sadakane et al. (2001) to be the lowest one found in HgMn stars.

A striking peculiarity of HR\,6000 is the lack of any
overabundance for heavy elements with Z$>$40, except for Xe and Hg.
The line spectrum of \ion{Xe}{ii} is similar to that we
observed in a preliminary analysis of UVES spectra of 46 Aql,
another \ion{Xe} rich star: while most of the lines lie at
the laboratory wavelength, some other \ion{Xe}{ii} lines seem to
be shifted by about $-$0.1\,\AA\ from the predicted position. 
The same behaviour can be observed also in HD\,175640, although to a less
extent, owing to its  lower \ion{Xe} overabundance.
Unfortunately, the small number of known transition probabilities 
and the ignorance of the \ion{Xe}{ii} isotopic structure due to the lack of atomic data put 
strong limitations in the study of this element in the spectra of the CP stars.

The large overabundance  of [+0.7] for iron generates a very rich line spectrum
in which numerous, still unclassified \ion{Fe}{ii} lines are observed.
A very large number
of lines probably due to \ion{Fe}{ii} remain unidentified.
The similar iron overabundance of 46 Aql ([+0.65]) gives rise to
an impressive  close resemblance between the spectra of the two stars. 
 
Abundances in the ultraviolet obtained from a re-analysis of the IUE spectra 
studied by Castelli et al. (1985) agree 
with abundances derived from the UVES spectra  for all the
elements, except for carbon and phosphorus. While the carbon identification
in the optical spectrum is rather questionable owing to the weakness of
the observed lines, the difference in the phosphorus abundance is similar 
to that yielded by the individual \ion{P}{ii} 
and \ion{P}{iii} lines.  

There are several signs of vertical abundance stratification in HR\,6000
for He, P, Mn, and Fe. The peculiar shape of the \ion{He}{i} profiles is 
discussed for the first time in this paper. The profiles have cores too  
intense as compared to the wings, a fact indicating strong vertical He 
abundance stratification (Dworetsky, 2004; Bohlender, 2005). 
Furthermore, the inferred
He abundance decreases with increasing wavelength ranging from an
underabundance of [-0.8] at 4000\,\AA\ to [-1.6] at 6000\,\AA. 

For \ion{Mn}{ii}, the abundance shortward of the 
Balmer discontinuity is larger by about 0.6\,dex than that longward of 
the Balmer discontinuity; for phosphorus, the \ion{P}{ii} abundance 
is larger by 0.2\,dex 
than the \ion{P}{iii} abundance, but this discrepancy lies within the 
mean square error of the average abundances;  for iron, the abundance from
\ion{Fe}{ii} high excitation lines is generally higher by about 0.2\,dex 
than that from \ion{Fe}{ii} low excitation lines.
   
Similar to other studied HgMn stars, also HR\,6000 shows emission lines.
The most numerous ones belong to \ion{Mn}{ii}, followed by those of \ion{Fe}{ii} and \ion{Cr}{ii}.
While the first two elements are overabundant in HR\,6000, chromium has solar abundance. 
The star exhibits also isotopic anomalies for Hg and Ca. 
In both cases the most heavy isotope is the predominant one.

\begin{acknowledgements}
We are very grateful to J.~F. Gonz\'alez for the re-reduction of UVES spectra using IRAF routines.
and to Dr. J.V. Smoker for the useful discussion regarding interstellar lines.
\end{acknowledgements}

\end{document}